\documentclass[useAMS,usenatbib]{mn2e}

\newcommand {\aap}{A\&A}
\newcommand {\apj}{ApJ}
\newcommand {\apjl}{ApJL}

\newcommand {\aj}{AJ}
\newcommand {\nat}{Nat}
\newcommand {\mnras}{MNRAS}

\usepackage{natbib,epsfig,graphicx,rotating,lscape,rotating}
\title[Transits in NGC 2362]{The Monitor project: the search for transits 
in the open cluster NGC 2362}
\author[A. Miller et al.]{Adam A. Miller$^{1,2}$\thanks{E-mail:
amiller@astro.berkeley.edu (AM)}, Jonathan Irwin$^{1,3}$, Suzanne 
Aigrain$^{4}$, Simon Hodgkin$^{1}$, Leslie Hebb$^{5}$.\\
$^{1}$Institute of Astronomy, University of Cambridge, Madingley Road, 
Cambridge, CB3 0HA, United Kingdom\\
$^{2}$Astronomy Department, University of California at Berkeley, Berkeley, CA 
94720-3411, USA\\
$^{3}$Harvard-Smithsonian Center for Astrophysics, 60 Garden Street, Cambridge,
MA 02138-1516, USA\\
$^{4}$Astrophysics Group, School of Physics, University of Exeter, Stocker 
Road, Exeter, EX4 4QL, United Kingdom\\
$^{5}$School of Physics and Astronomy, University of St. Andrews, North Haugh, 
St. Andrews, KY16 9SS, Scotland}
\begin{document}

\date{}

\pagerange{\pageref{firstpage}--\pageref{lastpage}} \pubyear{2008}

\maketitle

\label{firstpage}

\begin{abstract}

We present the results of a systematic search for transiting planets in a 
$\sim$5 Myr open cluster, NGC 2362. We observed $\sim$1200 candidate cluster 
members, of which $\sim$475 are believed to be genuine cluster members,
for a total of $\sim$100 hours. We identify 15 light curves with  
reductions in flux that pass all our detection criteria, and 6 of the 
candidates have 
occultation depths compatible with a planetary companion. The variability in 
these six light curves would require very large planets to reproduce the 
observed transit depth. If we 
assume that none of our candidates are in fact planets then we can place upper 
limits on the fraction of stars with hot Jupiters (HJs) in NGC 2362. We 
obtain 99\% confidence upper limits of 0.22 and 0.70 on the fraction of 
stars with HJs ($f_p$) for 1-3 and 3-10 day orbits, respectively, assuming 
all HJs 
have a planetary radius of 1.5$R_{\rm Jup}$. These upper limits represent 
observational constraints on the number of stars with HJs at an age 
$\la$10 Myr, when the vast majority of stars are thought to have lost their 
protoplanetary discs. Finally, we extend our results to the 
entire Monitor Project, a survey searching young, open clusters for planetary 
transits, and find that the survey as currently designed should be capable of 
placing upper limits on $f_p$ near the observed values of $f_p$ in 
the solar neighbourhood.

\end{abstract}


\section {Introduction}

\subsection {Transiting Planets Around a Young Star}

The detection of a transiting planet provides a wealth of information which 
cannot be matched by any other planetary detection method at the moment. There 
have been over 20 transiting planets detected thus far, and the list is growing 
rapidly\footnote{\tt http://obswww.unige.ch/$\sim$pont/TRANSITS.htm}. 
These systems provide the only means for directly 
measuring a planet's radius. They also provide a means for measuring the 
inclination of the orbital plane, which in turn removes the $\sin i$ ambiguity 
of a planet only detected via the radial velocity method. If the host star is 
bright enough, transiting planets also provide a means for
measuring their atmospheric composition when the stellar spectrum, in eclipse
(planet behind the star), is subtracted from the combined-light out of eclipse
spectrum of the star plus the planet (\citealt{richardson2007};
\citealt{grillmair2007}).
Additionally, very precise measurements of the time 
of transit can be used to search for and characterise the orbits of other, 
sometimes very low-mass, planets in the extrasolar system, because these 
other planets would slightly perturb the transiting planet's orbit 
(\citealt{agol2005}; \citealt{hm2005}).

Despite the recent success of many transit surveys, there has yet to be an 
observation of a transiting planet orbiting a PMS hydrogen burning star. In 
fact, there is a relative paucity of detected planets orbiting any stars at 
an age less than $\sim$100 Myr. Aside from the recent claim of the detection 
of a planet orbiting TW 
Hydrae \citep{setiawan2008}, we are unaware of any detections of a planet 
orbiting a $<$100 Myr hydrogen burning star.

Planets are believed to form in protoplanetary 
discs where formation and accretion, as well as migration, halt following 
the dispersal of the disc. \citet{hll2001} found that roughly 50\% of stars 
lose their discs by $\sim$3 Myr, while nearly all discs are dissipated by 
$\sim$10 Myr. This places a significant constraint on the formation time of a 
gaseous giant planet at $\la$10 Myr \citep{bl2002}. The discovery of a planet 
around a very young star, specifically $\sim$5 Myr, would  provide important 
constraints for planetary formation mechanisms, migration time scales and 
dynamical evolution, and their relation to disc lifetimes and clearing 
time scales \citep{bl2002}. A transiting planet at $\sim$5 Myr 
would lead to constraints on the planet mass and radius at an early age, i.e. 
very near the initial conditions for giant planet evolution, which are 
essentially unconstrained at the moment.

\subsection {The Cluster}

NGC 2362 is a well studied open cluster whose age ($\sim$5 Myr; 
\citealt{moitinho2001}, \citealt{delgado2006}), which coincides with the tail 
of the distribution of circumstellar disc lifetimes, and relatively moderate 
distance (1480 pc, $(m-M)_0$ = 10.85; \citealt{moitinho2001}) make it an 
ideal test bed for the study of stars and their environments during the early 
pre-main sequence (PMS).
For example, the detection of detached eclipsing binary (EB) 
stars in the cluster will provide measurements of each star's mass and radius 
without the use of a model. These measurements can, in turn, be used to 
constrain PMS evolution models.

Many recent surveys have been conducted to characterise the properties of NGC 
2362. The most relevant work includes the determination of fundamental 
cluster properties \citep{moitinho2001}, the determination of the 
circumstellar disc fraction from near-IR excess \citep{hll2001}, an 
H$\alpha$ emission survey to 
study T Tauri stars and disc accretion \citep{dahm2005}, 
and a study of primordial circumstellar discs using infrared excesses 
measured by the {\it Spitzer Space Telescope} \citep{dh2007}.

\subsection {The Survey}

We have completed a high cadence photometric monitoring survey of NGC 2362, 
with observations made using the Mosaic II imager on the 4m Blanco telescope 
at CTIO. There are three primary scientific goals of this survey: one, to 
discover low-mass EB systems which will allow us to simultaneously 
measure the mass and radius for each member of the system, two, to 
search for transiting planets orbiting PMS stars, and three, to 
characterise the rotation periods for low-mass members of the cluster 
\citep{irwin2007e}. 

Our observations of NGC 2362 were designed to be sensitive enough to detect a 
large planet ($R_{p}$ $\ga$ $1 R_{\rm Jup}$) in a `very hot-Jupiter' orbit, 
i.e. orbital period $\la$ 3 days \citep{aigrain2007}. Specifically, given 
the average observational 
cadence of $\sim$6 min during 18 nights spread over a year with optimal 
signal-to-noise we ought to be able to photometrically detect planets with 
$R_{p} > 1 R_{\rm Jup}$ around a 0.7 M$_{\sun}$ primary. For lower mass, and 
hence fainter, stars \citet{aigrain2007} predicted that we should be able to 
detect planets with $R_{p}$ $\ga$ $2 R_{\rm Jup}$ orbiting a 0.2 M$_{\sun}$ 
primary. This corresponds roughly with the spectroscopic limits from 8 m 
class telescopes. \citet{aigrain2007} shows that it would be possible to 
detect the radial velocity (RV) signal of a 1$M_{\rm Jup}$ planet orbiting a 
0.2 M$_{\sun}$ primary in a 1 day orbit down to $I \sim$ 18 using the UVES 
spectrograph on the Very Large Telescope. Initial 
simulations by \citet{aigrain2007} predicted that there would be 4.7 EBs 
and zero transiting planets in our data.

These observations are part of a larger photometric survey of 9 young 
($\la$ 200 Myr) open clusters covering a wide range of ages and metallicities 
(the Monitor project\footnote{\tt http://www.ast.cam.ac.uk/research/monitor/}; 
\citealt{hodgkin2006} and \citealt{aigrain2007}).

\subsection{Methods}

We developed an automated method to search for occultations in the light 
curves generated from our NGC 2362 observations.\footnote{From this point 
forward, when we refer to eclipses we specifically mean the eclipse of one 
star by another star or brown dwarf, while transits only 
refer to the case of a planet transiting a star. We use the term occultation 
in a more general sense to include both of these phenomena, but to also 
include all situations where the flux from a star has been reduced because 
some other body has passed in front of the stellar disc.} In particular, we
attempt to remove the significant spot-induced variability displayed by
many of our young and late-type targets. As
transits and eclipses are essentially the same from a detection
standpoint, and cannot always be distinguished from the light curve
alone, we detect both types of events among our candidates. We defer
the discussion of the EB candidates to a later paper.  Here we focus on our 
transit candidates, for which spectroscopic follow-up observations, to
confirm cluster membership and ascertain the source of the
occultations by measuring the mass of the occulting body, are underway.

We then perform extensive Monte Carlo simulations to evaluate our
sensitivity to planets of different radii and periods around stars of
different masses in the cluster, and use these simulations to place upper 
limits on the incidence of hot and very hot Jupiters (HJs and VHJs) at 5 Myr. 
These upper limits
represent constraints on HJ incidence at an age of less than
10 Myr, when any HJs should have recently finished migrating toward
their host star. They therefore constitute an important measurement for
constraining planetary formation and migration time-scales.

\subsection{Organisation of Paper}

The remainder of this paper is structured as follows: the data reduction 
process is discussed in Section 2, and our method of cluster membership selection 
is presented in Section 3. The technique used to search for occulting systems
and the resulting candidates are described in Section 4. The Monte Carlo 
simulations and the derived upper limits on the 
fraction of stars in NGC 2362 with short period planets, are discussed in 
Section 5. Here we also compare our sensitivity and results to other transit 
surveys that have targets stellar clusters. Finally, we present our conclusions
in Section 6.

\section {Observations and Data Reduction}

Photometric monitoring data were obtained using the CTIO 4m Blanco telescope, 
with the Mosaic-II imager, during 18 nights from February 2005 to January 
2006. This instrument provides a field of view of $\sim$36$\arcmin \times$ 
36$\arcmin$ (0.37 deg$^{2}$), using a mosaic of eight 2k $\times$ 4k pixel 
CCDs, at a scale of $\sim$0$\farcs$27 per pixel. A summary of our 
observations, including the number of frames per night as well as the start 
and finish time for each night, is given in Table~\ref{table:obs_summary}.

\begin {table}
  \centering
%\begin {center}
  \begin {tabular}{rcrr}
    \hline
    Night & Frames/night & Start & Finish \\
    \hline
    1 & 50 & 406.03573 & 406.29499 \\
    2 & 47 & 407.03439 & 407.28675 \\
    3 & 42 & 408.03985 & 408.30530 \\
    11 & 35 & 416.04713 & 416.24360 \\
    12 & 41 & 417.02969 & 417.24706 \\
    13 & 42 & 418.04069 & 418.24918 \\
    323 & 24 & 728.21098 & 728.35062 \\
    326 & 30 & 731.21755 & 731.36089 \\
    328 & 20 & 733.27306 & 733.36723 \\
    330 & 30 & 735.20998 & 735.36427 \\
    331 & 32 & 736.20901 & 736.36480 \\
    332 & 29 & 737.21096 & 737.36434 \\
    333 & 31 & 738.21689 & 738.36825 \\
    336 & 31 & 741.21493 & 741.36774 \\
    359 & 62 & 764.04498 & 764.33686 \\
    360 & 49 & 765.03671 & 765.33456 \\
    361 & 61 & 766.03490 & 766.32738 \\
    362 & 72 & 767.03074 & 767.33916 \\
    \hline
  \end {tabular}
\caption[Night is the night of observations relative to the first 
    night of observations. Start and Finish\ldots]{Summary of photometric 
        observations of NGC 2362. Night is the night of 
        observations relative to the first 
        night of observations. Start and Finish are the beginning and end of 
        the nightly observations, respectively, given in HJD - 2453000.5.}
\label {table:obs_summary}
%\end {center}
\end {table}

\begin {figure}
\begin {center}
% \vspace{302pt}
 \includegraphics [angle=270,width=84mm]{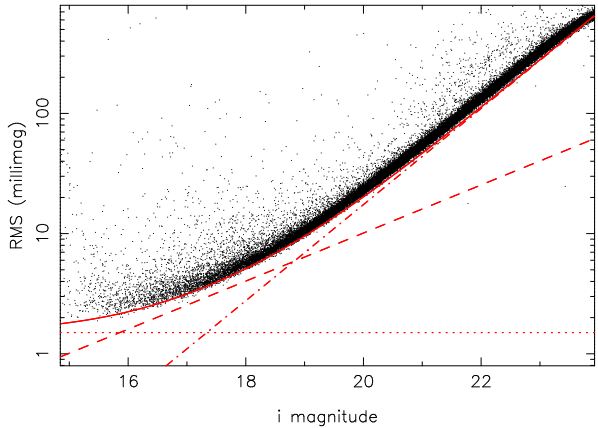}
  \caption{Plot of the rms scatter per data point for the entire series of 
    observations as a function of magnitude 
    for the $i$-band observations of NGC 2362. We include only unblended 
    objects with 
    stellar morphological classifications. The diagonal dashed line shows the 
    expected rms from Poisson noise in the object, the diagonal dash-dot line 
    shows the expected rms from sky noise in the photometric aperture, and the 
    dotted line shows an additional 1.5 mmag contribution added in quadrature 
    to account for systematic effects. The solid line shows the total 
    predicted rms from these effects. This plot shows the rms of our light 
    curves before they have been corrected for seeing.}
  \label{fig:rms_mag}
\end {center}
\end {figure}

Our observations consist of a series of 75 s $i$-band exposures with an 
average cadence of $\sim$6 min. We also obtained a few longer images 
in the $V$-band (3 $\times$ 600s) in photometric conditions for the production 
of a colour magnitude diagram (CMD). Our observations are sufficient to give 
1\% or better photometric precision per data point from saturation at $i 
\sim$ 15 down to $i \sim$ 19, as seen in Figure~\ref{fig:rms_mag}. This range 
corresponds to G through mid-M spectral types at the age and distance 
of NGC 2362.

For a full description of our data reduction procedure see 
\citet{irwin2007a}. Briefly, after we corrected for cross-talk between the 
detector readouts, we followed the standard CCD reduction scheme of bias 
correction, flat-fielding, and astrometric calibration before photometric 
calibration as described in \citet{il2001}. Following this we generated a {\it 
master catalogue} for the $i$-band filter by stacking 20 of the frames taken in
the best conditions (seeing, sky brightness and transparency) and running the 
source detection software on the stacked image. The resulting source positions 
were used to perform aperture photometry on all of the images, with the final 
result a time-series of differential photometry. 
We achieved a per data point photometric precision of $\sim$2-4 
mmag for the brightest objects, with RMS scatter $< 1\%$ for $i$ $\la$ 19 
(see Figure~\ref{fig:rms_mag}).

Our source detection software flags any objects detected as having overlapping
isophotes as likely blends. This information is used, in conjunction with 
a morphological image classification flag also generated by the pipeline 
software \citep{il2001}, to allow us to identify non-stellar or blended 
objects in the time-series photometry.

The CCD magnitudes were converted to the standard Johnson-Cousins system 
using regular observations of \citet{landolt1992} equatorial star fields in 
the usual way.

Light curves were extracted from $\sim$85 000 objects, 56 000 of which had 
stellar morphological classification, using our standard aperture photometry 
techniques \citep{irwin2007a}. We fit a 2D quadratic polynomial 
to the residuals in each frame (measured for each object as the difference in 
magnitude between the current frame and the median taken over all the frames) 
as a function of position, for each of the eight CCDs separately. We then 
removed this function to account for variations in transparency and 
differential atmospheric extinction across each frame. For a single CCD, the 
spatially varying part of the correction remains small, typically $\sim$0.02 
mag peak-to-peak.

As a last step there is a small correction applied to all the light curves 
for seeing-correlated effects. This was done by looking for seeing-correlated 
shifts in the light curve from its median magnitude. A simple quadratic 
polynomial was fit to the shift as a function of the full width half max of 
the stellar images on the corresponding frame. This fit was then subtracted 
from the light curve. Typically, this fit would reduce the rms of the light 
curve by $<$ 0.01 mag, however, for the cases that showed the strongest 
correlations with seeing the reductions in rms would be fairly significant, 
$\sim$0.1 mag.

For the production of deep CMDs, we stacked 20 $i$-band images taken in good 
seeing and photometric conditions. The long exposure $V$-band frames were 
stacked before running source detection and the accompanying astrometric and 
photometric calibration. The $V$-band sources were matched against sources in 
the $i$-band {\it master catalogue}, which then enabled us to produce a deep 
CMD. The limiting magnitudes, measured as the 
approximate magnitude at which our catalogues are 50\% complete were 
$V \simeq$ 24.4 and $i \simeq$ 23.6 \citep{irwin2007a}.  

\section {Selection of Low-Mass Candidate Members}

Before we could search for transits we had to identify low-mass 
cluster members. Lists of candidate members are available in the literature 
(\citealt{moitinho2001}, \citealt{dh2007}, \citealt{delgado2006}), however, 
in order to match the field of view of our survey, which is wider than  
previous surveys of NGC 2362, we elect to use a $V$ versus $V-I$ CMD for 
candidate membership selection. 

\subsection {The $V$ versus $V-I$ CMD}

\begin {figure}
\begin {center}
  \includegraphics [width=84mm]{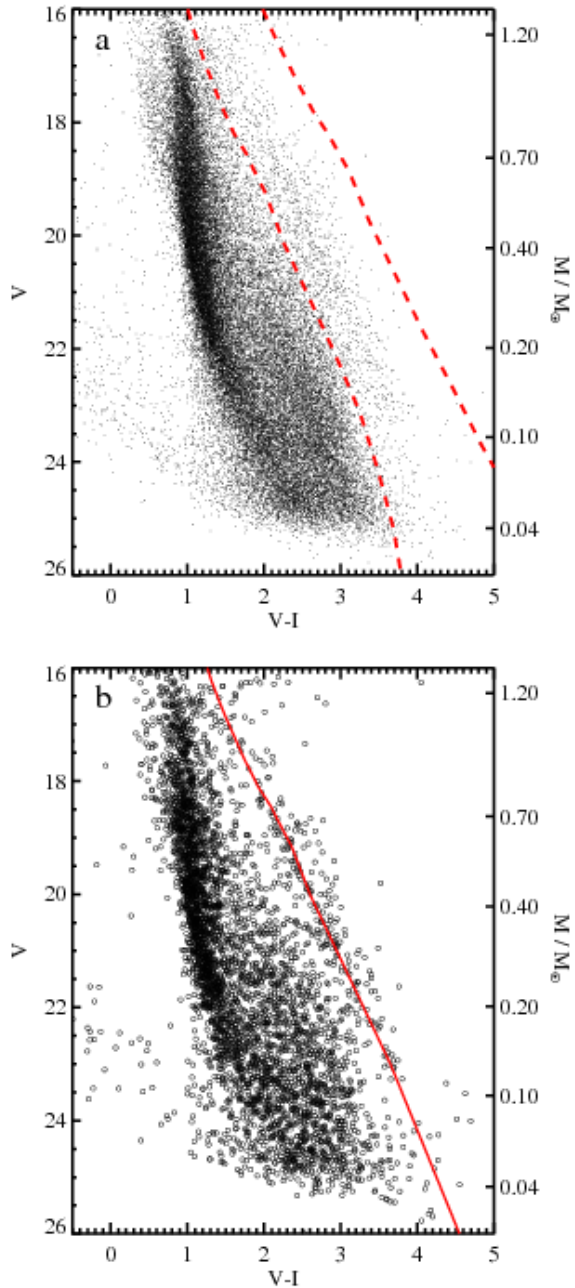}
  \caption{(a) $V$ versus $V-I$ CMD of NGC 2362 from stacked images for all 
    objects with stellar morphological classification. The cluster boundaries, 
    which define candidate membership, are shown as dashed lines 
    (all objects between the lines were selected). The cluster sequence is 
    clearly contaminated by objects in the galactic field. 
    (b) $V$ versus $V-I$ CMD of objects within $\sim$7 arcmin of $\tau$ CMa, 
    here defined as the centre of the cluster. The 7 arcmin cut was determined 
    by the spatial distribution of rotating stars, as described in the 
    text. The cluster sequence is clearly 
    visible and our empirical sequence is shown as the solid line.
    The mass scale in both plots is from the 5-Myr NextGen model isochrone 
    \citep{baraffe1998}, using our empirical 
    isochrone to convert the $V$ magnitudes to $I$ magnitudes, and 
    subsequently obtaining the masses from the $I$ magnitudes due to known 
    problems with the $V$ magnitudes of the models (see Section  3.1).}
  \label {fig-cmd}
\end {center}
\end {figure}

The $V$, $V-I$ CMD used for candidate membership selection is shown in 
the upper panel of Figure~\ref{fig-cmd}. The $V$ and $i$ measurements were 
converted to the standard Johnson-Cousins photometric system using colour 
equations from our observations of photometric standard stars 
(see Eqns. 1-3 from \citealt{irwin2007e}).

The cluster sequence is difficult to identify by eye, especially when compared 
to the CMDs of \citet{moitinho2001} and \citet{dahm2005}. Our survey covered a 
much larger portion of the sky (0.37 deg.$^{2}$ compared to $\sim$0.05 
deg.$^{2}$ for 
\citealt{moitinho2001} and $\sim$0.03 deg.$^{2}$ for \citealt{dahm2005}) and 
therefore suffers from greater field contamination. Thus, we used a 
second CMD including only objects within 7 arcmin of $\tau$ CMa, defined as 
the centre of the cluster, to identify 
the cluster sequence, shown in the lower panel of Figure~\ref{fig-cmd}. The 7 
arcmin radius was chosen based on a 2$\sigma$ cut from a Gaussian fit to the 
radial profile of rotating stars in the field of view.
Short period rotators, which are 
indicative of youth and therefore also cluster membership, were selected from 
the identifications by \citet{irwin2007e}. The presence of rotators at large 
($>$ 15 arcmin) distances from the centre of the cluster justifies the use 
of the entire field of view in our candidate selection process, despite the 
ensuing high contamination from field objects. We elected to do this because 
we did not want to miss any transiting planets, which are rare in 
both the cluster and the field.

We then followed the candidate selection method described in 
\citet{irwin2007e}. Briefly, we manually defined an empirical cluster sequence 
that follows the sequence visible in Figure~\ref{fig-cmd}b. All objects 
falling between two cuts defined by shifting the sequence right and left 
(perpendicular to the sequence itself) were then selected as candidate 
members. This led to the selection of 1813 candidate members over the full 
range from V=15.7 to 26.

To determine model masses and radii for our candidates, 
the $I$-band absolute magnitudes of the NextGen model are used, because these 
are less susceptible to a missing source of opacity, which creates a 
discrepancy between the models and 
observations in the $V - I$ colour for $T_{eff}$ $\la$ 3700 K (corresponding 
in this case to $V - I~\ga$ 2.5 \citealt{baraffe1998}). Therefore the $I$-band 
absolute magnitudes give the most robust estimates of mass and radius. Our 
adopted mass-radius-magnitude relation comes from the procedure described in 
\citet{aigrain2007}, which combines the NextGen isochrones of 
\citet{baraffe1998}, with the DUSTY isochrones of \citet{cbah2000}, and the 
COND isochrones of \citet{baraffe2003}. This adopted relation covers a mass 
range from 0.5M$_{\rm Jup}$ to 1.4M$_{\sun}$. For the 
few, $\sim$30, objects that are brighter than the limits of the 
NextGen models (i.e. $M_{I} \la$ 3.9, or $M_* > 1.4$M$_{\sun}$) we use the 
isochrones of \citet{siess2000} to determine the model masses and radii of 
those particular candidate cluster members.

\subsection {Contamination}

\citet{irwin2007e} estimate the level of contamination for the sample of 
candidate 
cluster members to be $\sim$65\%. They caution that their estimate is somewhat 
uncertain due to the need to use Galactic models. \citet{irwin2007e} also 
note that spectroscopic follow-up will be needed to make a more accurate 
contamination estimate. We note that when the range of magnitudes 
is restricted to those that we search for occultations (see Section 4) the 
contamination level is slightly reduced to $\sim$60\%.

\section {Occultation Detection}

While we are nominally searching for transits, our search procedure identifies 
any light curve with occulting events. Therefore, throughout this section we 
will discuss our search for occultations, which will inevitably yield a list 
of transit 
candidates. After achieving the necessary signal-to-noise, removing systematic 
trends, and obtaining sufficient coverage and sampling in the data, perhaps 
the most significant obstacle in any systematic search for occulting 
systems is the intrinsic variability present in many stars' light curves. 
Periodic variability, typically due to the rotation of spots on the surface 
of the star, is a particularly severe contaminant because it leads to regular 
reductions in the observed flux from the star, which is precisely the behaviour 
(i.e. occultations) we are trying to identify. Non-periodic variability is also 
significant in young stars: for instance, a star could change brightness 
following occultations by or interactions with  circumstellar material 
(see \citealt{bouvier2007} for a discussion of AA Tau, a young star whose 
photometric variability originates from interactions with its disc). 
\citet{dh2007} found an upper limit of $\sim$7\% for the fraction of stars 
with optically thick discs in NGC 2362. 
\citet{irwin2007e} note the difficulties in determining the fraction of 
cluster members that rotate, because the rotation sample has a lower 
contamination level than the remainder of the general candidate cluster 
members. Applying the same contamination estimates to both populations 
they estimate a conservative lower limit of $\sim$14\% for the fraction of 
cluster members which are rotating. \citet{irwin2007e} estimates that the 
actual fraction of cluster members which rotate is $\sim$40\% based on the 
high correlation between rotation and cluster membership. Clearly, there are 
more rotators than stars with optically thick discs, which is fortunate because
rotational variability is easier to filter than non-periodic variability. The 
effects of rotation lead to smooth variations in brightness. When 
this is the dominant source of variability it can be subtracted from the 
signal without introducing significant additional features into the 
light curve.

Before we began our search we removed from our sample a number of light curves 
which were flagged by the data reduction process \citep{irwin2007a}. 
Specifically, we removed any light curves that were flagged as saturated 
and any systems where more 
than 10\% of the data points belonged to low confidence regions in the 
standard size aperture as flagged by our reduction procedure 
\citep{irwin2007a}. We found a few stars which displayed behaviour consistent 
with saturation were not flagged in the original procedure. Thus, we elected 
to visually examine the remaining light curves by 
eye and flag those that were saturated. This led to the removal of an 
additional 75 objects. Objects that were flagged as blended were not 
excluded, however, because the reduction procedure of \citet{irwin2007a} did 
a sufficient job in correcting the effects of blending such that these 
objects could be searched for occultations along with the rest of the sample.

We also limited our search with a magnitude cut such that we only examine
objects brighter than $I$ = 19. We place these cuts based on the limitations 
of spectroscopic follow-up: the cause of occultations must be 
confirmed with RV measurements (for a full description of the spectroscopic 
limitations of this study see \citealt{aigrain2007}). Briefly, 
\citet{aigrain2007} found that with existing multi-object spectrographs on 
8 m class telescopes it would be possible to reach RV precision of $\sim$
2 km/s down to $I \sim$ 18. Our cut of $I$ = 19 is thus 
conservative, yet we point out that \citet{sahu2006} were able to measure 
RV variations to $\sim$1 km/s for an object with $I$ = 18.75 using the 
UVES echelle spectrograph at the 8-m Very Large Telescope, while 
\citet{weldrake2007c} were able to detect a $K$ = 114 m/s planet in a 4-day 
orbit around a K star of $V$ = 17.4. We note that any 
candidates fainter than $\sim$18 in $I$ will be extremely 
difficult to follow-up in order to confirm the origins 
of the occulting behaviour found in the light curve.

Following the removal of these objects there remained 1180 candidate cluster 
members to be searched for occultations.

\subsection {Variability Filtering}

The difficulty with attempting to remove the intrinsic stellar variability 
from a light curve is that many filtering methods will remove or affect the 
occultation signal 
as well as the photometric signal coming from just the star alone. Therefore, 
the challenge lies in developing a method that successfully disentangles the 
occultation signal from that of the host star.

\begin {figure}
\begin {center}
  \includegraphics [width=84mm]{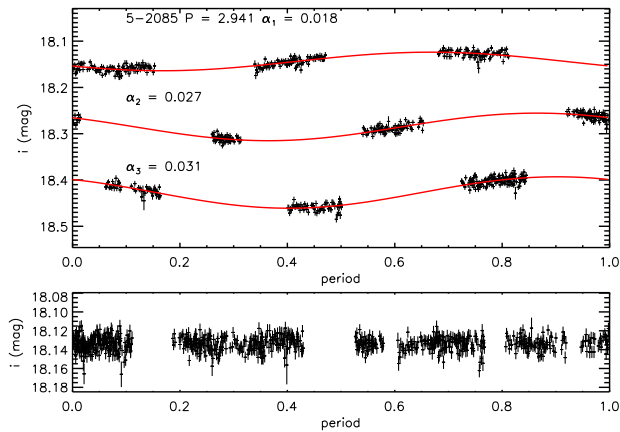}
  \caption{Example rotation fit to star 5-2085. Shown is the light curve and 
    best fit folded over the rotation period (top) and the period-folded 
    sine-subtracted light curve (bottom). The three different observing 
    windows, as described in the text, are shown with 0.15 mag offsets in the 
    top panel. It is clear to see that both the amplitude and phase of the 
    signal from this star change with time. $P$ is the best fit period in days 
    while $\alpha_i$ is the best fit amplitude in magnitudes for each of the 
    observing windows.}
  \label{fig:rotation-fit}
\end {center}
\end {figure}

Because we find rotation to be the dominant source of variability in the 
candidate members of NGC 2362, we focused on removing this source of 
variability. After $k\sigma$ clipping outlying data points in the light curves,
we selected rotating stars in our sample by performing a least-squares 
sine-fit to the time series $m(t)$ (in magnitudes) of every candidate cluster 
member using 

\begin {equation}
m(t) = m_{dc} + \alpha~sin({2\pi}t/P + \phi),
\label{eqn1}
\end {equation}
where $m_{dc}$ is the mean light curve level, $\alpha$ the amplitude, $\phi$ 
the phase and $P$ the period of rotation. For the fits $m_{dc}$, $\alpha$, 
and $\phi$ are free parameters at each value of $P$ over a grid of equal 
logarithmically spaced steps in period from 0.1-181 d (corresponding to half 
the time between our first and last observations). Our fits adopted a single 
period, 
but we allowed the phase and amplitude to change following any gaps in our 
observations of more than three weeks. We allowed these changes because the 
size and location of star spots can evolve very rapidly over these time 
scales in young stars. We fix the period, however, because we would not expect 
a significant change in the angular momentum of the star in the course of 
a single year. The output of this procedure is a `least-squares 
periodogram,' and the best-fitting period is the one with the lowest reduced 
$\chi^{2}$. An example of this fitting procedure is shown in 
Figure~\ref{fig:rotation-fit}, where the data have been folded on the best fit 
period. From the figure it is easy to see the change in amplitude and phase 
following long gaps in our observations.

\begin {figure}
  \begin {center}
  \includegraphics [width=84mm]{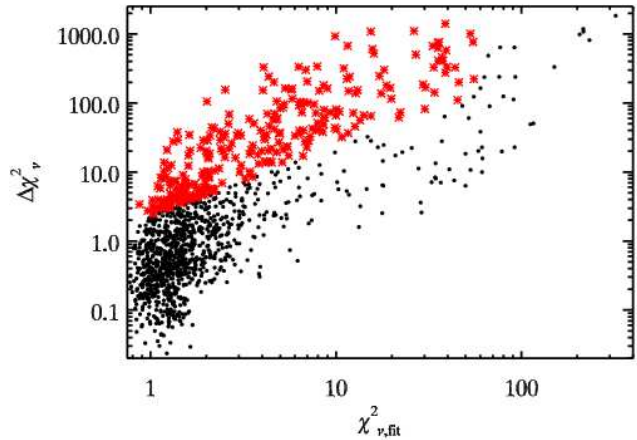}
  \caption{$\Delta\chi_{\nu}^{2}$ as a function $\chi^2_{\nu,{\rm fit}}$ for all 
    candidate cluster members in our sample. The stars highlight the systems 
    that were selected as rotators.}
  \label{fig:rot_select}
  \end {center}
\end {figure}

We measured the reduced $\chi^{2}$ in our light curves before 
($\chi^2_{\nu,{\rm flat}}$) and after ($\chi^2_{\nu,{\rm fit}}$) we 
subtracted the best sine-fit and selected rotators based on the change in 
reduced $\chi^{2}$:
\begin {equation}
  \Delta\chi_{\nu}^{2}/\chi^2_{\nu,{\rm flat}} > 0.7,
  \label{eqn2}
\end {equation}
where $\chi^2_{\nu,{\rm flat}}$ is the reduced $\chi^{2}$ of the original light 
curve with respect to a constant model, and $\Delta\chi_{\nu}^{2}$ is the 
change in reduced $\chi^{2}$ following the fit. We also required that 
$\chi^2_{\nu,{\rm fit}} <$ 60 for an object to be classified as a rotator. We 
acknowledge that a reduced $\chi^2$ of 60 is quite large, however, we find that 
these systems ($\chi^2_{\nu,{\rm fit}} \sim$ 60) display periodic variability. 
The large value of $\chi^2$ is the result of extremely high signal-to-noise 
data and slight departures from the modelled behaviour of Eqn.~\ref{eqn1}. The 
upper limit for $\chi^2_{\nu,{\rm fit}}$ was selected because we found that 
Eqn.~\ref{eqn1} was a poor model for all objects above this threshold. 
Figure~\ref{fig:rot_select} shows $\Delta\chi_{\nu}^{2}$ plotted against 
$\chi^2_{\nu,{\rm fit}}$, and highlights the systems which were selected as 
rotators. 

\begin {figure}
\begin {center}
  \includegraphics [width=84mm]{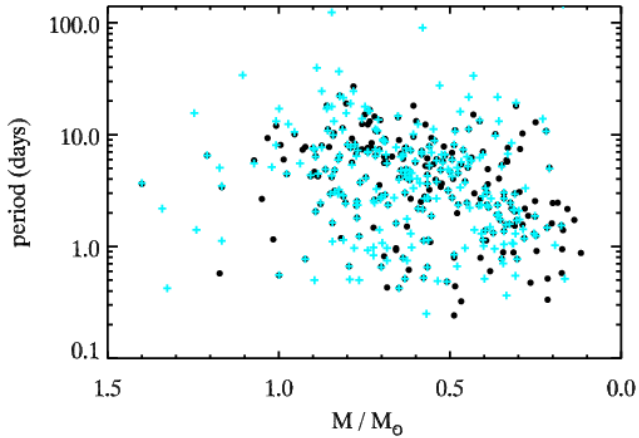}
  \caption{Rotation period versus mass for the candidate cluster members 
    classified as rotators. The blue crosses show those objects classified 
    as rotators in this paper. The large black points show rotators as 
    identified by \citet{irwin2007e}. \citet{irwin2007e} only use a portion 
    of the data, so they are not sensitive to periods greater than 
    $\sim$24 days. We note that not every cross represents a rotator, 
    especially the excess of objects at $\sim$0.5 and $\sim$1 d (due 
    to night-edge effects). Other discrepancies between the two samples are 
    likely the result of different windows of observation, such that in some 
    stars spot activity was likely low during the subset of the data that 
    \citet{irwin2007e} examined.}
  \label {fig:pmd}
  \end {center}
\end {figure}

The empirically determined cutoff in Eqn.~\ref{eqn2} is more stringent than 
the initial cut used in the Monitor rotation papers (\citealt{irwin2006}, 
\citealt{irwin2007b}, and \citealt{irwin2007e}). Those papers employ visual 
inspection of each light curve to remove any objects without clear periodic 
variability from their sample. Therefore, their samples are more complete. 
We wish to avoid visual inspection, however, in order to remove human 
interaction from our selection procedure. As noted by \citet{burke2006}, a 
clear set of detection criteria that do not rely on human input are extremely 
important for establishing the actual sensitivity of a survey to planetary 
transits using Monte Carlo simulations. Therefore we chose not to add a step 
with visual inspection. 
We do find that there is generally good agreement between our sample and the 
one in \citet{irwin2007e}, except for the slowest 
rotators. \citet{irwin2007e} use data from only a single observing season, so 
they are not sensitive to periods $\ga$ 24 days. We also note that our 
classification scheme leads to 
some objects, whose variability is not the result of star spots, to be 
misclassified as rotators, as can be seen by the slight build up 
around 0.5 and 1 days in the period distribution of Figure~\ref{fig:pmd}. 
This build up is caused by low amplitude night edge effects.
For the purposes of this study this less than perfect classification 
scheme is acceptable because the subtraction of the sine-fits was designed to 
remove periodic variability of a non-eclipsing nature.

Our goal was to identify any objects which clearly displayed periodic 
variation of a relatively long temporal signature while excluding those with 
high-frequency events where the reduced $\chi^{2}$ would be significantly 
improved by a sine-fit (i.e. objects with multiple eclipses). We wanted to 
test that our classification scheme did not identify transits as rotation. 
To do this we took a 
sample of our flattest light curves, selected for their low $\chi^{2}$ 
($\chi^2_{\nu,{\rm flat}} <$ 1.5) and low dispersion in median flux for each 
observing season, and inserted transits according to the formalism of 
\citet{ma2002} using the limb-darkening coefficients of \citet{claret2000}. 
The planet radius and orbital period were chosen randomly over 
the same grid as that used in the simulations in Section 5. The orbital 
inclination and epoch were chosen such that we would have observed at least a 
portion of one transit. We then ran the rotation test described above and we 
found these objects to be misclassified as rotators in $<$ 0.1\% of our test 
cases. This extremely low rate of false classification allowed us to proceed 
with confidence that we were not removing occultation signals from our light 
curves. We were not worried about misclassifying EBs because we found that the 
removal of a sine curve from an EB light curve left large residuals which were 
detected by our occultation search algorithm.

Following this procedure we detected rotation in 268 stars, or roughly 23\% 
of the candidate cluster members in our sample. In the objects where rotation 
was detected we subtracted the best sine-fit from the light curve and searched 
for occultations in the same manner used for non-rotators described below. 

\citet{ai2004} propose a number of alternative methods for filtering 
intrinsic stellar variability without removing the signal from an occultation. 
We attempted to use their least-squares filtering method and non-linear 
filtering method but we found these removed the signal from even the deepest 
eclipses. \citet{ai2004} designed these methods for space based occultation 
surveys, and caution that their procedures are likely only valid for 
observations which are continuous on time scales $\gg$ a typical occultation. 
Unfortunately this is rarely, if ever, the case with ground based surveys and 
we confirm their initial forewarnings. 

Lastly, we also attempted to model the variations using the formal spot model 
of \citet{dorren1987}. We found this method to be far too computationally 
intensive for an automated procedure, while the least-squares sine-fit 
serves as a good proxy to the full spot model.

\subsection {Noise Properties}

Before we began our search for occultations we examined our light curves for 
the presence of correlated, or red, noise. Ground based surveys have been 
shown to suffer from red noise (see \citealt{pzq2006} for a very detailed 
discussion, and the typical red noise levels in several existing surveys), 
which invariably makes it more difficult to detect occulting objects. This 
correlated noise means that the uncertainty in data binned over $n$ points 
decreases slower than in the uncorrelated, or white, noise case, where the 
uncertainty in binned data is $\propto$ 1/$\sqrt{n}$. 

Our search for occultations is based on a Box Least Squares (BLS) fit, where 
the box represents a short time scale periodic decrease in the mean flux from 
the star. Therefore correlated noise on the same time scale as an occultation 
can lead to a large detection statistic for every light curve, even those 
which are spurious. A large detection statistic in every light curve 
necessitates an extremely large detection threshold, meaning that only the 
most significant occultations are followed up, while shallow occultations or 
light curves with only a few occultation data points are not detected. 
\citet{pzq2006} showed that it is possible to modify the standard white noise 
detection statistic to account for correlated noise, however, thereby 
eliminating many of the spurious candidates. 

\begin {figure}
\begin {center}
  \includegraphics [width=84mm]{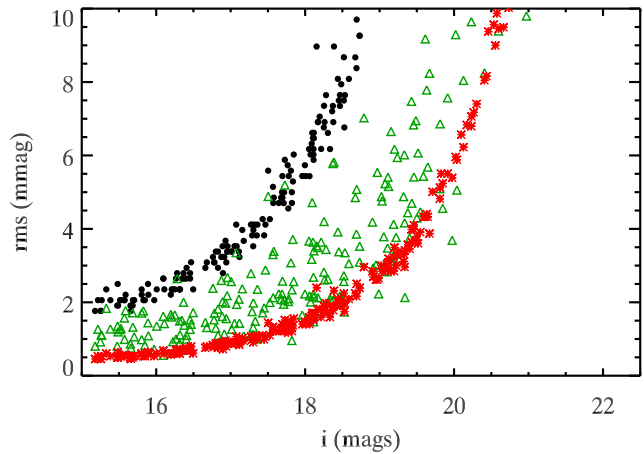}
  \caption{RMS as a function of magnitude for a subset of candidate cluster 
    members in NGC 2362. The filled circles represent the RMS scatter per data 
    point, the triangles the RMS for 15-point averages, and the stars 
    represent the expected values of the 15-point averages assuming white 
    noise. The triangles lie well above the stars indicating the presence of 
    red noise over $\sim$2.5 hour time scales. For the brightest stars this 
    appears to be the dominant effect in the noise.}
  \label {fig:correlated-noise}
\end {center}
\end {figure}

We characterise the presence of red noise in our survey according to the 
method outlined in \citet{pzq2006} by examining the flattest light curves 
(low reduced $\chi^2$ with respect to a constant model) in our sample. 
These objects, which exhibit little variability, should be dominated by  
noise. Figure~\ref{fig:correlated-noise} shows the 
RMS scatter for individual points as a function of magnitude as well as the 
RMS in 15-adjacent-point averages (which for the sampling of our data 
corresponds to roughly 2.5 hours, a typical time scale for transiting hot 
Jupiters), compared to the expected value of the RMS in 15-adjacent-point 
averages in the presence of white noise. It is clear to see that the expected 
1/$\sqrt{n}$ decrease in the noise does not apply to most of our light curves. 
Therefore, an assumption of white noise is unfounded, and any attempt to 
detect occultations must take this red noise into consideration.

For our survey we find correlated noise at a level of $\sim$1 mmag for the 
brightest stars in our sample. This is on a similar scale to the best ground 
based surveys discussed in \citet{pzq2006}. 

\subsection {Search for Occultations}

Following the removal of periodic variability we searched all the candidate 
cluster members for occultations using a refined version of the BLS 
algorithm \citep{kzm2002} as designed by \citet{ai2004}. 
\citet{ai2004} show that the best fit model reduces to finding the inverse 
variance weighted mean of the observed occultation data points. They test the 
significance of a detection using the familiar detection statistic $S$: 
%%%%
\begin {equation}
S^2 = \left(\sum_{i \in I} \frac{d_i}{\sigma_i^2}\right)^2 \left(\sum_{i \in I} \frac{1}{\sigma_i^2}\right)^{-1},
\label {eqn3}
\end {equation}
where the sum includes all in-occultation data points $i$, $d_i$ is the 
difference in flux between the $i$th data point and the mean flux of the 
entire light curve, and $\sigma_i$ is the uncertainty in the $i$th flux 
measurement.
\citet{ai2004} also show that $S^2$ is equal to the difference in $\chi^2$ 
between a flat model and the best fit occultation model of the data. 
Detections with large $S$ are typically considered the best candidates for 
spectroscopic follow-up. \citet{ai2004} only considers the case of white 
noise, however, which can lead to many spurious 
candidates. We discuss the significance of a detection in the presence of red 
noise below.

\begin {figure}
\begin {center}
  \includegraphics [width=84mm]{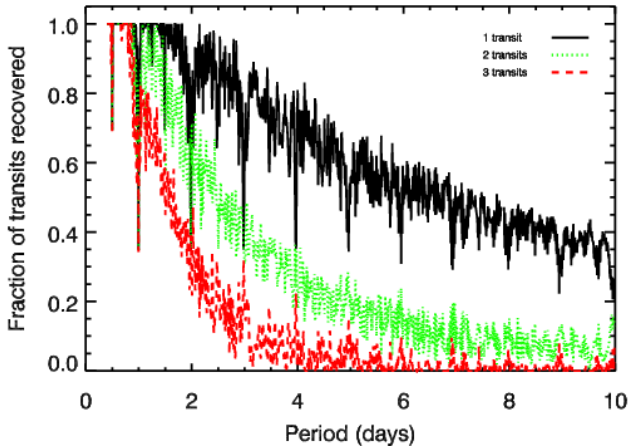}
  \caption{Fraction of transiting systems recovered as a function of orbital period. The 
    results for 1 (top, solid curve), 2 (middle, dotted curve), and 3 (bottom, 
    dashed curve) detected transits are shown.}
  \label{fig:recovery}
  \end {center}
\end {figure}

We restricted our search to occultations that happen in a period range of 0.4 
- 10 days. The upper limit was chosen because \citet{aigrain2007} show that  
the time sampling of our observations, combined with the geometric probability 
of a transit, is insensitive to planets with orbital periods greater than 
$\sim$10 days. The 
lower limit was selected below the typical boundary of $\sim$1 day because 
we wanted to search for systems with extremely short periods, and because we 
wanted to test our sensitivity at these short periods in the simulations 
described in Section 5. It is important to remember, however, that there are 
no confirmed planet detections to date with periods less than 1 day.
(\citealt{sahu2006} found 5 planetary candidates with 
periods less than a day, and as low as 0.42 days, however, they remain 
unconfirmed because they are too faint for follow-up spectroscopy.) We show 
our transiting system recovery fraction as a function of period and the total 
number of 
required transits for a positive detection in Figure~\ref{fig:recovery}. The 
figure gives the probability of 1, 2, or 3 transits being present in the 
data as a function of orbital period. Transits are considered to be present if 
there are data within the range of phases $\phi < 0.1W/P$ or $\phi > 1 - 
0.1W/P$ where $W$ is the expected transit width and $P$ is the orbital period, 
i.e. a transit is present if we observe some portion of the central 20\% of 
the transit. 
Given that $\ge$ 3 detections are needed to accurately the determine the 
period of a system, Figure~\ref{fig:recovery} shows that we can only hope to 
recover reliable periods for systems with orbital periods $\la$ 2 days.

While searching each light curve for the best-fitting occultation model, we also 
fit for the best brightening model (defined as an increase in brightness, 
rather than decrease as is the case for an actual occultation) as described in 
\citet{burke2006}. This secondary fit has little impact on the numerical 
efficiency of the fit, because the algorithm simultaneously searches for the 
greatest reduction in $\chi^2$ in terms of both brightening and dimming. We 
have no reason to believe that objects with correlated noise on the time scale 
of an occultation or non-periodic variable objects not selected by the method 
described in Section 4.1 preferentially show correlated decreases, rather than 
increases, in flux. On the contrary, one would expect that for these cases 
both an occultation and brightening could provide good models to the data. 
Therefore, we use the ratio of the improvement in $\chi^2$ 
for an occultation model to a brightening model to discriminate against 
non-eclipsing objects by requiring $\Delta\chi^2/\Delta\chi^2_- >$ 3, where 
$\Delta\chi^2$ is the improvement in $\chi^2$ using an occultation model and 
$\Delta\chi^2_-$ is the improvement in $\chi^2$ relative to a brightening model.

\begin {figure}
\begin {center}
  \includegraphics [width=84mm]{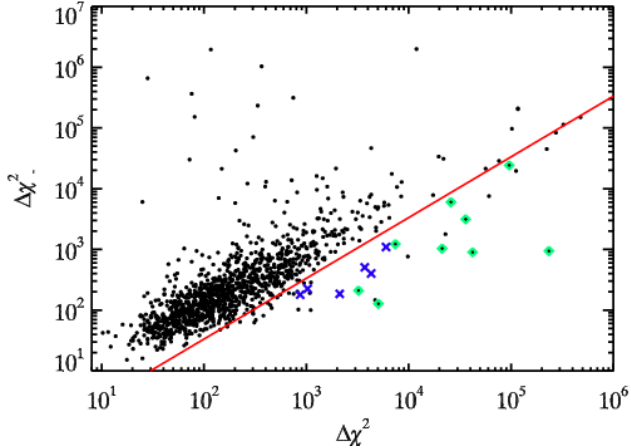}
  \caption{The improvement in $\chi^2$ with respect to a brightening model 
    ($\Delta\chi^2_-$) as a function of the improvement in $\chi^2$ with 
    respect to an occultation model ($\Delta\chi^2$) for every searched light 
    curve. The solid line corresponds to $\Delta\chi^2/\Delta\chi^2_- =$ 3. 
    We require points to lie below the line to be considered occultation 
    candidates. The green diamonds indicate light curves that pass all of 
    our selection criteria. The blue X's are a subset of the green diamonds. 
    In addition to passing all our selection criteria, the blue X's have 
    observed occultation depths that may be compatible with a planetary 
    companion (see Section4.4).}
  \label{fig:brightening}
  \end {center}
\end {figure}

Figure~\ref{fig:brightening} shows a plot of $\Delta\chi^2_-$ against 
$\Delta\chi^2$ for every light curve in our sample. The solid line shows the 
boundary of our requirement $\Delta\chi^2/\Delta\chi^2_- >$ 3. Objects below 
the line pass the selection criteria and are modelled significantly better by 
an occultation than by brightening. 

We employed the method of \citet{pzq2006} to determine the significance of a 
detection for each light curve in the presence of red noise. They show that 
the detection statistic, $S_{\rm red}$ in a light curve with red noise can be 
found without a fit to the individual white and red noise components. 
We briefly summarise the procedure here: first we found the best 
fit solution assuming white noise (this is equivalent to maximising $S$), then 
we masked points in-occultation. We then calculated the mean flux over a 
sliding interval equal to the duration of the detected occultation, where the 
sliding steps are smaller than the interval 
between flux measurements. We grouped those flux measurements into bins based 
on the number of data points in the sliding interval, and we calculated the 
variance of the flux measurements in each bin. These variances give an 
estimate 
to what \citet{pzq2006} call the $V(n)$ function, where $V(1)$ equals the 
variance in bins with only one point, $V(2)$ equals the variance in bins with 
two points, and so on. The detection statistic can be measured with $V(n)$ 
using Eqn.~7 from \citet{pzq2006}, reproduced here:
\begin {equation}
S_{\rm red}^2 = d^2\frac{n^2}{\sum_{k=1}^{N_{tr}}n_k^2V(n_k)},
\label {eqn5}
\end {equation}
where $d$ is the depth, $n$ is the total number of data points in-occultation, 
the sum is over all occultations, and $n_k$ is the number of points in the 
$k$th occultation.

\begin {figure}
\begin {center}
  \includegraphics [width=84mm]{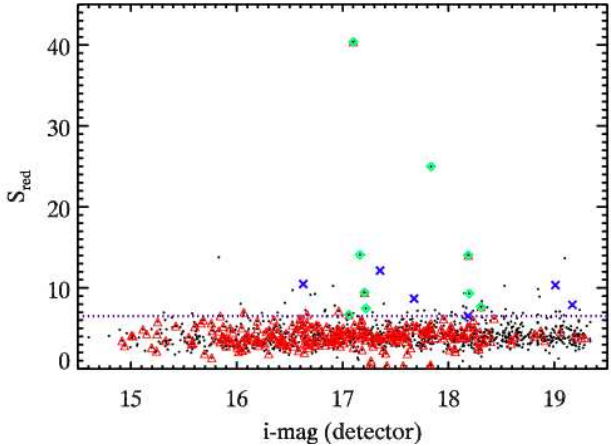}
  \caption{Detection statistic accounting for red noise as a function of 
    CCD $i$-band magnitude for all candidate cluster members brighter 
    than $I$ = 19 (black dots). Objects with large values of $S_{\rm red}$, 
    which occur over the full magnitude range of our observations, indicate 
    our best occultation candidates. The red triangles show stars classified 
    as rotators. The dashed horizontal line shows our adopted threshold in 
    $S_{\rm red}$. Stars above the line meet the criteria of the threshold. The 
    green diamonds and blue X's are the same as Figure~\ref{fig:brightening}.}
  \label{fig:Sred-mag}
\end {center}
\end {figure}

\begin {figure}
\begin {center}
  \includegraphics [width=84mm]{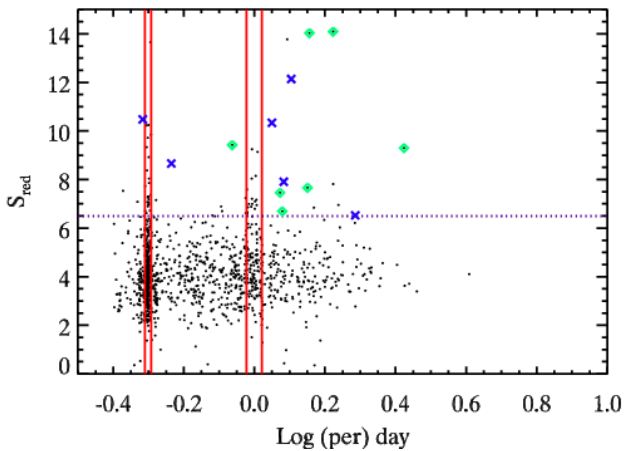}
  \caption{$S_{\rm red}$  as a function of best fit occultation period for all 
    candidate cluster members in our sample. 
    The vertical lines show regions of detected periods that 
    are excluded to remove false positives, as described in the text. The 
    dashed horizontal line shows our adopted threshold in $S_{\rm red}$. Stars 
    above the line meet the criteria of the threshold. We note that the 
    scale of $S_{\rm red}$ has been reduced, which excludes the two candidate 
    occultation systems at ($i$, $S_{\rm red}$) $\approx$ (17.1, 40) and 
    (17.6,25) in Figure~\ref{fig:Sred-mag}. The green diamonds and blue X's 
    are the same as Figure~\ref{fig:brightening}.}
  \label{fig:Sred-per}
\end{center}
\end{figure}

After measuring $S^2$ for each light curve, we determined the corresponding 
value of $S_{\rm red}$. We show $S_{\rm red}$ as a function of detector $i$-band 
magnitude in Figure~\ref{fig:Sred-mag} for every object in our sample. From 
the figure it can be seen that we found strong detections ($S_{\rm red} \ga 6$) 
over the full magnitude range of our observations. The fact that there is 
no bulk trend in $S_{\rm red}$ as a function of magnitude shows that 
$S_{\rm red}$ 
takes into account any residual correlated noise, which is typically magnitude 
dependent. The triangles show objects classified as rotators following the 
procedure in Section 4.1. Figure~\ref{fig:Sred-mag} indicates that our filtering 
method is successful, as the vast majority of the rotators have detection 
statistics consistent with non-eclipsing light curves. 
A plot of $S_{\rm red}$ as a function of best fit period can be 
seen in Figure~\ref{fig:Sred-per}. We find a large over-density of points at a 
period of 0.5 days. A histogram of the best fit period for every object shows 
a high frequency at 0.5 and 1 days. Visual inspection of these light curves 
shows a systematic effect where slight changes in the median flux ($\sim 
0.005$ mag) from one observing season to another can be fit extremely well 
with periods of $\sim$0.5 or $\sim$1 days. We are not entirely certain 
why the flux changes from season to season for this subset of objects, but we 
wish to eliminate these false 
positives. Therefore, we do not consider any light curves with a best fit 
period within $1 \pm 0.05$ and $0.5 \pm 0.01$ days as detections. The vertical 
lines in Figure~\ref{fig:Sred-per} indicate these regions of discarded periods.

\subsubsection {Detection Threshold}

Following their examination of the effects of correlated noise on the ability 
of transit surveys to detect planets, \citet{pzq2006} use simulated light 
curves to show that false positives rarely have values of $S_{\rm red}$ larger than 
7. They then argue that the adopted detection threshold will vary from survey 
to survey, but that it will typically be in the range 7-9. 

Following from this, we chose to adopt an inclusive detection threshold of 
$S_{\rm det}$ = 6.5. Figures~\ref{fig:Sred-mag} and~\ref{fig:Sred-per} 
show that very few light curves have $S_{\rm red} > S_{\rm det}$, while a 
histogram of $S_{\rm red}$ for all objects within the allowed range of best 
fit occultation periods shows very few counts when $S_{\rm red} >$ 6.5. We 
acknowledge that the selection of this cut will likely increase our false 
positive rate relative to other transit surveys, however, in our case this 
may not prove to be entirely detrimental. We investigate the choice of our 
detection threshold in Section 5, and find that our Monte Carlo simulations 
corroborate our adopted detection thresholds.

There are a few circumstances that make our 
situation fairly unique. The first is that the size of the cluster matches 
the FLAMES field of view \citep{aigrain2007}. The second is the high 
scientific value of any transit candidates which turn out to be cluster EBs. 
Therefore, we can afford to adopt a relatively low threshold and accept more 
false positives than a typical survey, because (1) we can follow-up many of 
these candidates at the same time and (2) the detection of any occulting body 
in the cluster provides an important discovery. In Section 5, we use a series of 
Monte Carlo simulations to estimate our sensitivity and false positive rate 
based on our selection of $S_{\rm det}$.

\subsection {Occultation Candidates}

15 out of the 1180 candidate cluster members pass all of our 
selection criteria as occultation candidates. Following depth considerations, 
planetary transits can be ruled out for all but six of the candidates. 
For the cases where transits cannot be ruled out very large planets, 
$>$ 1.5$R_{\rm Jup}$, would be needed to explain the observed occultation 
depth.

We estimate the occultation depth for each of the 15 candidates in order 
to exclude any systems that could not be a dwarf-planet system from our list 
of transit candidates. The depth is measured by eye in the following way: 
the minimum flux during occultation is subtracted from the out of occultation 
flux on the same night. This results in an estimate of the depth in 
magnitudes. We measure the occultation depth by eye because intrinsic stellar  
variability makes the best fit depth from our occultation search unreliable. 
We do not rule out planets in any candidate where the observed occultation 
depth could be explained by a planet with $R_p \le 3R_{\rm Jup}$. This leaves 
six candidates as possible transits, while the other nine systems remain EB 
candidates. We defer a discussion of the EB candidates to a later paper. In 
Table~\ref{table:eclipses2} we include the observed occultation depths and 
corresponding minimum planet radii necessary to explain the occultation.  We 
note that the uncertainty in our estimates of the necessary planetary radius, 
which relies on stellar evolution models and a visual measurement of the 
occultation depth, is large. Table~\ref{table:eclipses2} also contains the 
occultation parameters of each of our transit candidates. 
$\Delta\chi^2/\Delta\chi^2_-$ and $S_{\rm red}$ are the detection statistics 
discussed in Section 4.3. 

\begin {table}%{lllrrrrrrlrrrrr}
\begin {center}
%\centering
\begin {tabular}{lrrrr}
   \hline
   \hline
   star & $\Delta\chi^2/$ & $S_{\rm red}$ & $\delta$ & $R_p$ \\
     & $\Delta\chi^2_-$ & & (mag) & ($R_{\rm Jup}$) \\
   \hline
3-10048 & 10.76 & 12.1 &  0.05 & 2.49 \\
3-3739 & 7.35 & 10.5 &  0.03 & 2.37 \\
3-7559 & 4.85 & 6.5 & 0.05 & 2.07 \\
5-6469 & 4.58 & 8.7 & 0.03 & 1.71 \\
6-9484 & 5.48 & 7.9 & 0.17 & 2.63 \\
7-4723 & 11.44 & 10.3 & 0.15 & 2.62 \\
   \hline
  \end {tabular}
\caption[Candidate occulting systems in NGC 2362.
  Star is the object identification number from this work.\ldots]{Fit 
    parameters for candidate transiting systems in NGC 2362.
    Star is the object identification number from this work.  
     $\delta$ is the 
    occultation depth, measured by eye, and $R_p$ is the corresponding 
    planet radius necessary to account for a transit of that depth 
    assuming a central transit. 
 \label{table:eclipses2}}
\end {center}
\end {table}
%\end {landscape}

Table~\ref{table:eclipses} summarises the properties of the six stars where 
we cannot rule out the possibility of a transit. We include the survey 
identification number, J2000 coordinates, optical photometry taken at CTIO, 
$NIR$ photometry 
from 2MASS where available, as well as the model mass and radius for each 
candidate from our adopted magnitude-mass-radius relation (see Section 3). These 
estimates of the stellar parameters assume that the observed flux 
is from a single star, and that there is no occulting body which contributes 
to the total brightness of the system. If these systems are EBs, then these 
values serve as upper limits to the mass and radius of the primary star. 

\begin {table*}%{lllrrrrrrlrrrrr}
\begin {tabular}{lrrrrrrrrrr}
   \hline
   \hline
   star & $\alpha$ & $\delta$ & $V$ & $V - R$ & $V - I$ & $J$ & $J-H$ & $H-K$ & M$_*$ & $R_*$ \\
     & (J2000) & (J2000) &(mag) & (mag) & (mag) & (mag) & (mag) & (mag) & (M$_{\sun}$) & (R$_{\sun}$) \\
   \hline
3-10048 & 7:18:40.95 & -24:53:00.9 & 19.93 & 1.39 & 2.85 & 15.43 & 0.68 & 0.15 & 0.53 & 1.21 \\
3-3739 & 7:17:51.68& -24:53:03.3 & 19.12 & 1.30 & 2.76 & 14.95 & 0.65 & 0.34 & 0.73 & 1.41 \\
3-7559 & 7:18:21.73& -24:55:04.2 & 20.67 & 1.32 & 2.76 & 16.49 & 1.17 & 0.23 & 0.32 & 0.95 \\
5-6469 & 7:19:34.23 & -25:11:28.6 & 20.44 & 1.43 & 3.06 & 15.69 & 0.64 & 0.22 & 0.44 & 1.11 \\
6-9484 & 7:19:56.21 & -25:02:22.2 & 22.13 & 1.58 & 3.28 & ... & ... & ... & 0.18 & 0.71  \\
7-4723 & 7:19:24.32 & -24:54:05.4 & 21.99 & 1.62 & 3.30 & ... & ... & ... & 0.20 & 0.75 \\
   \hline
  \end {tabular}
\caption[Candidate occulting systems in NGC 2362.
  Star is the object identification number from this work. $V$, 
 $V - R$, and $V - I$\ldots]{Candidate transiting systems in NGC 2362.
  Star is the object identification number from this work. $\alpha$ and 
  $\delta$ are the RA and DEC, respecitively. $V$, 
 $V - R$, and $V - I$ are optical photometry measurements from this work. 
 $J$, $J -H$ and $H-K$ are $IR$ measurement from 2MASS, where available. M$_*$ 
 and $R_*$ are the model mass and radius, respectively, using our adopted 
 magnitude-mass-radius relation, and assuming the candidate is being occulted 
 an object or material which does not contribute to the total brightness of 
 the  system. \label{table:eclipses}}
\end {table*}

\begin {figure}
\begin {center}
  \includegraphics [angle=90,width=84mm]{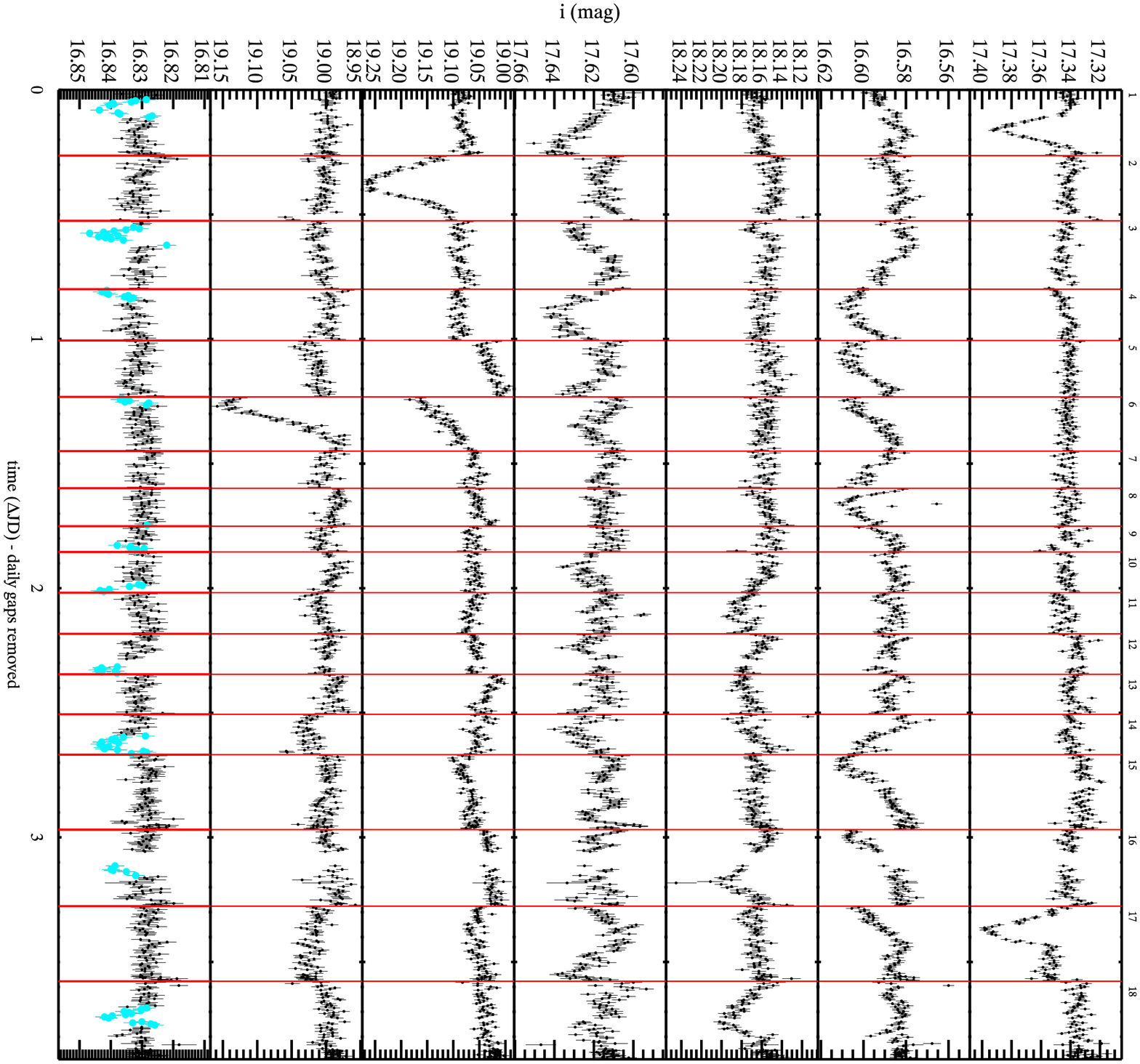}
  \caption{Full light curves for the six stars where we could not rule out the 
  planetary hypothesis. From top to bottom: 3-10048, 3-3739, 3-7559, 5-6469, 
  6-9484, 7-4723. The bottom panel shows a light curve with an inserted 
  transit (see Section 5). The transits have been highlighted relative to the 
  out-of-transit data.  Note that the simulated transit is typically shorter 
  in duration and shallower than the events shown above. Gaps in our 
  observations have been removed and are indicated 
  by the vertical lines. For reference the relative night of observation is 
  listed above the corresponding portion of the light curve.}
  \label{fig:pp}
\end {center}
\end {figure}

In Figure~\ref{fig:pp} we show the full light curves for each of our transit 
candidates, as well as a simulated transit included as a reference. We elect to 
show the entire light curve because in most cases, 
with the exception of a few, we do not detect enough occultations to 
accurately determine the period of the system. We show the observed light 
curves before they have been filtered for variability, because 
out of occultation variability is indicative of youth, and hence cluster 
membership. 
Note that for the model transit the duration is shorter, and the depth is 
shallower than for our candidate systems. This indicates that there is a good 
chance that our transit candidates are in fact EB systems.

We note that for each case multi-epoch mid- to high-resolution 
spectroscopy is going to be needed for RV measurements in order to (1) 
determine cluster membership via the systemic velocity and (2) confirm whether 
the observed reductions in flux are the result of an 
occulting planet or dwarf. Toward these two goals, we will observe NGC 2362 
with VLT/FLAMES. In most of 
these cases additional photometric observations will be needed in order to 
provide better orbital phase coverage as well as additional detections of 
the single or partial occultations already observed. 

\section {Results}

\subsection{Simulations}

We report here on the Monte Carlo simulations designed to characterise the 
sensitivity of our survey to transits and estimate the false positive 
rate. We do this by inserting transits into the raw data, correcting 
for correlations with seeing (see Section 3), and searching the 
resulting light curves using the same procedure described in Section 4. 
We consider all simulated light curves to be positive detections if prior 
to the injection of a transit $\Delta\chi^2_-/\Delta\chi^2_+ <$ 3 and/or 
$S_{\rm red} <$ 6.5, whereas following the insertion of a planet 
$\Delta\chi^2_-/\Delta\chi^2_+ >$ 3 and $S_{\rm red} >$ 6.5. Saturated stars 
and stars fainter than $I$ = 19 were not included 
in the simulations. We only insert transits into 1171 of the 1180 light curves 
we initially searched for occultations in Section 4. The 9 stars where we 
ruled out the planetary hypothesis based on the depth of the occultations 
were excluded from the simulations, because they have large reductions in 
flux that would result in positive detections regardless of the shape 
or size of the inserted transit signal. 

We use the simulations to 
characterise the survey's sensitivity to HJs. We define the sensitivity, 
${\cal S}$, as 
\begin{equation}
{\cal S} = \frac{n_d}{N_{sim}},
\end{equation}
where $n_d$ is the total number of planetary systems that we detect and 
$N_{sim}$ is the total number of simulated planetary systems. This represents a 
measure of our ability to detect planets over the entire range of possible 
inclinations, not just those planets that transit their host star. 
From ${\cal S}$, we can determine the expected number of planet detections, 
following some assumptions about the fraction of stars with planets.

These simulations accurately address the actual signal to noise in our 
data, the real cadence and observing windows in our data, and, perhaps most 
importantly, the intrinsic variability present in the light curve of each 
individual star. This represents an improvement to the simulations  
in \citet{aigrain2007}, which assume a signal to noise based on the average 
noise properties of non-variable light curves and ignore the intrinsic 
variability of individual stars. 

When testing the sensitivity of a transit survey it is important 
to inject realistic transits into the data \citep{burke2006}. Therefore, we 
adopt the formalism of \citet{ma2002}, which provides 
a method for calculating analytic transits given the ratio of planet and 
star radii as well as the geometry of the orbit. Specifically, we use the 
quadratic limb-darkening model presented in Section 4 of \citet{ma2002}. We 
determine the limb-darkening coefficients from the tables of 
\citet{claret2000}, and the stellar properties come from the isochrones of the 
NextGen, DUSTY, and COND models, as described in Section 3. We 
acknowledge the perils of assuming the validity of the very models we are 
trying to test, however, without the use of some model to relate absolute 
magnitude to stellar mass and radius the following simulations would be 
impossible. 

For our simulations we assume circular orbits. Given the relatively low 
eccentricities of all discovered extrasolar planets orbiting within 0.1 AU, 
this is a reasonable assumption. For planets with semi-major axis less than 
0.1 AU the median eccentricity is 0.04, with a mean eccentricity $<e>$ = 0.08, 
and only two planets with $e >$ 0.3.\footnote{\tt http://exoplanet.eu/catalog.php}

Each simulation was run using a uniform distribution in log$_{10}$ period from 
0.4 to 10 days, while the epoch was selected randomly from a uniform 
distribution of orbital phase. Finally we selected an orbital inclination from 
a uniform distribution in $\sin i$. Transits only occur when the 
planet passes in front of the stellar disc, i.e. only in cases where 
\begin {equation}
(R_* + R_p)/a \ge \cos i, 
\label{eqn:6.1}
\end {equation} 
%%%%%
where $R_*$ is the stellar radius, $R_p$ is the planet radius, and $a$ and $i$ 
are the orbital semi-major axis and inclination, respectively. Therefore, we 
could choose to select to only simulate the cases where the inserted planet 
transits the stellar disc. We choose to select $i$ from the full range of 
orbital inclinations, however, because we want to use the simulations to test 
our sensitivity to all planets, regardless of whether the planet transits 
its host star or not. This choice has a negligible effect on the cost of the 
total computing time. In the cases where the planet does not transit (i.e. 
$i$ does not satisfy Eqn.~\ref{eqn:6.1}), or the phase coverage is such 
that we do not observe any transits, we do not need to run the search 
procedure as the light curve remains unchanged and the results of the 
occultation search are the same as those already obtained during the 
initial search procedure described in section Section 4.

We inserted planets of radii $1R_{\rm Jup}$ and $1.5R_{\rm Jup}$ into the light 
curves. We arrived at the 
lower planetary radius because $\sim 1R_{\rm Jup}$ is roughly the lower limit 
of radii we are sensitive to following the preliminary simulations of 
\citet{aigrain2007}. We chose the larger radius based on the evolutionary 
models of \citet{burrows1997} for extrasolar giant planets. According to 
\citet{burrows1997} at an age of 5 Myr planets in the mass range 0.7$M_{\rm Jup}$ 
to 2$M_{\rm Jup}$ have radius $R_p \sim 1.5R_{\rm Jup}$, while the largest radius 
occurs for a Saturn mass planet (0.3$M_{\rm Jup}$) with $R_p \sim 1.7R_{\rm Jup}$.
\footnote{\tt http://zenith.as.arizona.edu/$\sim$burrows/} 

For every star included in the simulations we independently inserted 300 
planets with $R_p$ = 1$R_{\rm Jup}$ and 300 planets with $R_p$ = 1.5$R_{\rm Jup}$. 
Specifically, for each iteration of the simulations we would: (1) add a planet 
to the system with given $R_p$, $i$, and $a$, which comes from the adopted 
stellar mass and randomly selected period, (2) check to see if 
Eqn.~\ref{eqn:6.1} is satisfied, if it is then (3) insert the transit into 
the light curve, and (4) check to see if the randomly chosen epoch results in 
any observed transits. In total this corresponds to the insertion 
of 708 000 simulated planetary systems of which we searched $\sim$
101 500 and $\sim$105 500 systems with observed transits for the 
inserted $1R_{\rm Jup}$ and $1.5R_{\rm Jup}$ planets, respectively.

\subsubsection {False Positives}

False positives cannot be identified by a light curve alone. RV measurements, 
which can determine the planetary nature and cluster membership of any 
candidates, are needed to fully characterise the fraction of light curves that 
are in fact false positives. \citet{brown2003} calculates the expected number 
of false positives due to stellar companions for transit surveys of the field 
star population. An analysis similar to that of \citet{brown2003} adapted to 
the stellar population in NGC 2362, which is different from field stars, would 
be desirable, but is considered beyond the scope of this paper.

\subsubsection {Sensitivity as a Function of Detection Threshold}

We now evaluate our selected thresholds based on the results of our 
simulations. In Table~\ref{table:threshold} we summarise the sensitivity 
and maximum false positive fraction for a number of different thresholds 
in $S_{\rm red}$ 
and $\Delta\chi^2/\Delta\chi^2_-$ for both 1$R_{\rm Jup}$ and 1.5$R_{\rm Jup}$.
As noted above, we cannot determine the number of false positives from the 
light curves alone. If we assume that there are no transiting planets in 
our data, then we can, however, determine an upper bound to the number of 
false positives. This upper bound is determined by assuming that all systems 
which pass the detection criteria are 
considered false positives:
\begin{equation}
{\cal FP} = \frac{n_{\rm cand}}{1171},
\end{equation}
where ${\cal FP}$ is the upper bound on the number of false positives, 
$n_{\rm cand}$ is the number of systems that pass all the selection criteria, 
and 1171 is the total number of searched systems.

The quoted values of ${\cal S}$ assume that every star has a HJ, which means 
the number of detections relative to the number of false positives cannot 
be found by dividing ${\cal S}$ by ${\cal FP}$. To find this ratio 
${\cal S}$ would have to be multiplied by the fraction of stars with a HJ, 
or $\sim$1\% (\citealt{gaudi2005}; we caution that this result is limited to 
the solar neighbourhood and may not apply to a cluster at the distance of 
NGC 2362, where the formation environment may be significantly different 
than that in the solar neighbourhood).

\begin {table}%{lllrrrrrrlrrrrr}
\centering
\begin {tabular}{rcccccc}
   \hline
   \hline
 & $\Delta\chi^2/$ & \multicolumn{2}{c}{$1.0R_{\rm Jup}$} & &\multicolumn{2}{c}{$1.5R_{\rm Jup}$} \\   
\cline{3-4} \cline{6-7} 
$S_{\rm red}$ & $\Delta\chi^2_-$ & ${\cal S}$ & ${\cal FP}$ && ${\cal S}$ & ${\cal FP}$ \\
   \hline
 4.5 &  2.0 & 0.069 & 0.025  && 0.135 & 0.025  \\
 5.0 &  2.0 & 0.066 & 0.022  && 0.132 & 0.022  \\
 5.5 &  2.0 & 0.062 & 0.020  && 0.128 & 0.020  \\
 6.0 &  2.5 & 0.048 & 0.011  && 0.117 & 0.011  \\
 6.5 &  3.0 & 0.038 & 0.005  && 0.107 & 0.005  \\
 8.0 &  3.0 & 0.029 & 0.003  && 0.095 & 0.003  \\
10.0 &  5.0 & 0.012 & 0.002  && 0.072 & 0.002  \\
15.0 & 10.0 & 0.001 & 0.000  && 0.038 & 0.000  \\
   \hline
  \end {tabular}
\caption{Sensitivity, ${\cal S}$, assuming a logarithmic distribution in 
orbital periods, and the upper bound to the fraction of false positives 
(${\cal FP}$) as a function of planet radius. Note that ${\cal S}$ and 
${\cal FP}$ are not directly 
comparable because the quoted values of ${\cal S}$ assume every star has a HJ.
\label {table:threshold}}
\end {table}

What Table~\ref{table:threshold} clearly shows is that there is very little 
improvement in the sensitivity when the $S_{\rm red}$ threshold is lowered 
below 5.5, while at the same time ${\cal FP}$ grows relatively 
quickly. At the same time, to obtain ${\cal FP} <$ 0.1\% would 
require the selection of very large detection thresholds. Given the relatively 
small number of cluster stars in this survey, such a large threshold would be 
inadvisable. Reducing the thresholds from these large values to $S_{\rm red} 
\sim$ 6 and $\Delta\chi^2/\Delta\chi^2_-~\sim$ 3 results in the largest gains 
in ${\cal S}$ relative to the increase in ${\cal FP}$. We 
arrive at our final thresholds of $S_{\rm red}$ = 6.5 and $\Delta\chi^2/
\Delta\chi^2_-$ = 3, because reducing each of those thresholds by 0.5 creates 
additional occultation candidates whose variability is clearly not the 
result of eclipses or transits.

If we assume that 1\% of stars have a HJ then the values quoted in 
Table~\ref{table:threshold} are somewhat discouraging. This would mean that, 
given our adopted thresholds, for a large survey in order to detect 1 planet 
at 1$R_{\rm Jup}$ and 1.5$R_{\rm Jup}$ we would have $\sim$13 and $\sim$5 false 
positives, respectively. This is somewhat abated 
 considering this cluster lies within a singe FLAMES 
field of view, meaning that the telescope time needed for spectroscopic 
follow-up is essentially independent of the number of candidates. 

\subsubsection {Sensitivity as a Function of Stellar Mass}

As noted previously, the presence of a planet cannot be inferred from a light 
curve alone. Given that consistent RV variations are needed to 
confirm a planet, it would be very useful to understand our sensitivity as a 
function of stellar mass for stars in the cluster. Our ability 
to detect planets in NGC 2362 via RV variations is strongly dependent on the 
mass of the host star \citep{aigrain2007}. Assuming a Jupiter mass planet, the 
greatest signal in RV variations is going to occur for the lowest mass stars, 
however, these stars are also going to be the faintest in the cluster meaning 
it will be difficult to obtain spectra with sufficient signal-to-noise to 
detect the RV variations. While moving to higher mass stars will dramatically 
improve the signal-to-noise in a single spectrum it will cause a significant 
reduction in the RV amplitude, again assuming a Jupiter mass companion. This 
will make it difficult to detect a planet. Thus, an understanding 
of the sensitivity as a function of stellar mass becomes extremely important 
when determining which systems to follow-up for this cluster and for the 
Monitor project as a whole.

\begin {table}%{lllrrrrrrlrrrrr}
\centering
\begin {tabular}{ccccc}
   \hline
   \hline
mass & radius &\multicolumn{1}{c}{$1.0R_{\rm Jup}$}&&\multicolumn{1}{c}{$1.5R_{\rm Jup}$} \\   
\cline{3-3} \cline{5-5} 
 (M$_{\sun}$) & (R$_{\sun}$) & ${\cal S}$ & & ${\cal S}$ \\
   \hline
$<$ 0.3 & $<$ 0.92 & 0.048 && 0.114 \\
0.3 - 0.5 & 0.92 - 1.18 & 0.038 && 0.104 \\
0.5 - 0.7 & 1.18 - 1.38 & 0.027 && 0.102 \\
$>$ 0.7 & $>$ 1.38 & 0.037 && 0.109 \\
   \hline
  \end {tabular}
\caption{Sensitivity as a function of stellar mass and radius. The mass and 
  radius are determined from our adopted magnitude-mass-radius relation as 
  described in the text. ${\cal S}$ is the sensitivity for each range of 
  masses.
\label{table:mass}}
\end {table}

In Table~\ref{table:mass} we summarise the sensitivity to HJs 
as a function of stellar mass. We also show the corresponding radii from our 
adopted magnitude-mass-radius relation. Somewhat surprisingly, 
Table~\ref{table:mass} shows that we are more likely to recover a planet 
around a smaller star. This is surprising because the stars with the lowest 
pre-transit rms are all at the bright (and hence higher mass) end of the 
cluster. This implies that the effect of reduced stellar radii is more 
important than a low rms when trying to detect transits. This confirms
the initial findings of \citet{ap2007}, who examined the detectability of 
transits in a hypothetical cluster. \citet{ap2007} found that when 
red noise is considered it becomes more difficult to detect planets around the 
brighter cluster members even though they have a smaller rms. Our results 
suggest that small stellar radii may be the most important factor in detecting 
transits given that our survey is most sensitive to planets orbiting stars 
with mass $< 0.3$M$_{\sun}$ and 
radius $< 0.92$R$_{\sun}$. 

\subsection {Expected Number of Hot Jupiter Detections}

Using the Monte Carlo simulations described above we estimate the expected 
number of detectable short period planets in NGC 2362. Many factors 
must be accounted for in order to estimate the number of expected detections, 
including: the frequency of HJs, the total number of candidate cluster members 
and the contamination from field stars, the geometric transit probability as 
a function 
of period, and our recovery rate as a function of period. We note that the 
relatively small number of cluster members in NGC 2362 means that a null 
result in NGC 2362 does not carry much significance. 

We begin by assuming that the frequency of large, short period planets is the 
same as that found in the solar neighbourhood. \citet{gaudi2005} identify 
two empirically defined populations of planets with periods $\la$ 10 days. 
Following from their work we assume the frequency of hot Jupiters, 
$O_{\rm pl}$, with periods of 3-10 days, to be 1\%. Strictly speaking, the 
estimates from \citet{gaudi2005} include only planets with periods up to 9 
days. We extend their estimate to 10 days without loss of generality. The 
frequency of HJs in the other population with periods $<$ 3 days, which 
\citet{gaudi2005} refer to as VHJs, is $\sim$0.15\%. This 
is then multiplied by the number of candidate cluster members, $N_*$, and the 
probability of membership, $P_{memb}$, to arrive at the expected number of HJs 
in our data set. The number of HJs is not randomly distributed in orbital 
period. The actual distribution of orbital periods is currently ill 
constrained, though it likely depends 
on the physical parameters of the system in which the planet is formed. In 
order to calculate the number of expected detections we must make some 
assumption about the period distribution. Therefore, we assume a uniform 
logarithmic distribution in orbital period. We then separate the expected 
number of HJs into equal size bins in log(period) space and calculate the 
number of planets in each period bin $i$, which we then sum to arrive at 
\begin {equation}
N_{\rm det} = \sum_{i = m}^n N_{\rm pl}(per_i),
\label{eqn:N-detect}
\end {equation}
with
\begin {equation}
N_{\rm pl}(per_i) = N_*P_{memb}O_{\rm pl}f_{\rm pl_i}{\cal S}_i
\end {equation}
where $N_{\rm det}$ is the total number of detectable HJs present in the 
data set, $N_*P_{memb}$ is the number of cluster members we observed, 
$O_{\rm pl}$ is the frequency of HJs in the solar neighbourhood, $f_{\rm pl_i}$ is an 
estimate of the 
fraction of the total number of HJs in bin $i$ based on a logarithmic 
distribution of orbital periods, and ${\cal S}_i$ is our sensitivity to 
planets in the $i$th period bin. Our final assumption is that there are no HJs 
with periods less than 1 day, because to date there have been no RV confirmed 
planets found with periods $<$ 1 day.

The resulting expected distribution of HJs is shown in 
Table~\ref{table:N-detect}. We find that we would only expect to detect 0.19 
HJs in NGC 2362, which is consistent with a null detection. The values quoted 
in Table~\ref{table:N-detect} assume that HJs orbiting young stars have bloated 
radii of 1.5$R_{\rm Jup}$. If instead we assume that these planets all have radii 
of 1.0$R_{\rm Jup}$ then the sensitivity is reduced and the expected number of 
detectable HJs in our data is reduced to 0.038. Under either 
circumstance, it is clear that we would not expect to find any HJs in our 
data. 

\begin {table}%{lllrrrrrrlrrrrr}
\centering
\begin {tabular}{cccc}
   \hline
   \hline
Period & & &\\   
(days) & $F_{\rm pl_i}$ & ${\cal S}_i$ & $N_{\rm pl}(Per)$\\
   \hline
0.4-1 &  0.00 & 0.228 & 0.000 \\
1-3 &  0.71 & 0.094 & 0.066 \\
3-10 &  4.72 & 0.027 & 0.126 \\
   \hline
  \end {tabular}
\caption{Number of expected detectable, transiting planets in NGC 2362. We have 
  separated the results into 3 different populations based on orbital period. 
  $F_{\rm pl_i}$ is the expected number of HJs in our data in  
  the corresponding period range. ${\cal S}_i$ is the sensitivity to 1.5$R_{\rm Jup}$ 
  HJs, as determined by our Monte Carlo simulations. $N_{\rm pl}(Per)$ is the 
  expected number of transit detections at each period.
\label{table:N-detect}}
\end {table}

\subsubsection {Upper Limits on the Planetary Fraction in NGC 2362}

Given our null detection (assuming that none of our candidates are planetary 
transits, which has yet to be confirmed) we can place upper limits on the 
fraction of stars in NGC 2362 with a HJ. We do this by assuming that all stars 
in the cluster have a HJ, and then calculate the corresponding number of 
expected detections based on the results of our Monte Carlo simulations. 
Following from there we can evaluate the statistical significance of a null 
result given the number of expected detections. 

Based on RV surveys of the solar neighbourhood, we know that not every star 
has a HJ \citep{gaudi2005}. The actual number of stars with detectable HJs will
be a Poisson distribution with mean $\mu$ = $f_pN_{\rm det}$, where $f_p$ is 
the actual fraction of stars with HJs and $N_{\rm det}$ is the expected number 
of detections assuming all stars have a HJ. For a Poisson distribution the 
probability of $n$ detections given a mean $\mu$ is
\begin {equation}
P(n; \mu) = \frac{e^{-\mu}\mu^n}{n!}.
\label{eqn:poisson}
\end {equation} 
%%%%%%%%%%%%%%%%%%
For our null result $n$ = 0, which when we substitute into 
Equation~\ref{eqn:poisson} yields
\begin {equation}
P(0; f_pN_{\rm det}) = e^{-f_pN_{\rm det}}.
\label{eqn:probability}
\end {equation} 
%%%%%%%%%%%%%%%%%
To obtain an upper limit on $f_p$ at significance $\alpha$ we require 
that 
\begin {equation}
\alpha \ge P(0; f_pN_{\rm det}).
\label{eqn:upper-limit}
\end {equation} 
%%%%%%%%%%%%%%%%
When we substitute Equation~\ref{eqn:probability} into 
Equation~\ref{eqn:upper-limit} we arrive at 
\begin {equation}
f_p \le \frac{- \ln \alpha}{N_{\rm det}},
\label{eqn:ul-significance}
\end {equation}
%%%%%%%%%%%%%%%%%
which allows us to place an upper limit on $f_p$ at any significance $\alpha$ 
(or confidence level 1-$\alpha$). For example, if one expects to detect 3 
planets, then there is a 5\% chance of detecting zero planets from Poisson 
statistics alone (Eqn.~\ref{eqn:probability}). Equivalently, to convert a null 
detection to a statistical result at 95\% confidence or greater, the 
expectation value of the number of planets detected should be 3 or greater.

In Table~\ref{table:upper-limit} we show the derived upper limits on $f_p$ 
for $\alpha$ = 0.05 and 0.01 (corresponding to confidence levels of 95\% and 
99\%, respectively) for HJs with $R_p$ = 1.0$R_{\rm Jup}$ and 
1.5$R_{\rm Jup}$. We quote upper limits for two different period ranges:  
1-3 d, and 3-10 d. In order to calculate the upper limits in each  
period bin we assume that every star has one planet of radius $R_p$. We then 
determine the number of stars with a HJ in a given bin, assuming a uniform 
logarithmic distribution, and multiply this by the sensitivity, ${\cal S}_i$, 
in the given bin to arrive at $N_{\rm det}$. The most significant upper limits 
we find are at $\alpha$ = 0.01, where if HJs at $\sim$5 Myr are about 
1.5$R_{\rm Jup}$ then the upper limit on short (1-3 days) period HJs is 22\% 
while the upper limit on HJs with periods between 3-10 days is 70\%. These 
limits provide strong evidence against the 'survival of the lucky few' 
scenario, where most planets migrate into their host star and only those that 
form just prior to the dispersal of the disc survive, as discussed in 
\citet{aigrain2007}. We note that if HJs at this age tend to be closer to 
1$R_{\rm Jup}$ we cannot place reliable constraints on $f_p$.

\begin {table}%{lllrrrrrrlrrrrr}
\centering
\begin {tabular}{crccc}
   \hline
   \hline
$R_p$ & $N_{\rm det}$ & Period range & U. L. on $f_p$ & U. L. on $f_p$ \\   
($R_{\rm Jup}$) &  & (days) & at $\alpha$ = 0.05 & at $\alpha$ = 0.01 \\
   \hline
1.0 &  5.75 & 1.0 - 3.0 & 0.521 & 0.801 \\
1.0 &  1.03 & 3.0 - 10.0 & 1.000* & 1.000* \\ %2.898 & 4.456 \\
1.5 & 21.15 & 1.0 - 3.0 & 0.142 & 0.218 \\
1.5 &  6.61 & 3.0 - 10.0 & 0.453 & 0.697 \\
   \hline
  \end {tabular}
\caption{Upper limits on the fraction of stars with HJs. $R_p$ is the planet 
  radius. $N_{\rm det}$ is the expected number of detections if every star has 
  a HJ. $f_p$ is the fraction of stars with HJs in the given period bin, and 
  $\alpha$ is the significance level of the upper limit on $f_p$. {\it *}For 
  the case of 3 - 10 day 1$R_{\rm Jup}$ planets we cannot place upper limits 
  below every star in the cluster having a HJ in this period range.
\label{table:upper-limit}}
\end {table}

\subsection {Discussion}

Given the number of cluster stars ($\sim$475), the number of hours observing 
($\sim$100), and the faintness of many of the cluster members it is unlikely 
that we would have detected a planet in NGC 2362. The failure to detect a 
planet is unsurprising based on our simulations, and the simulations of 
\citet{aigrain2007}, which predicted zero detectable planets in our data for 
this cluster (again, we are assuming that none of our transit candidates 
are in fact transiting planets, which remains to be confirmed). In fact, the 
detection of a planet would have been more inconsistent with the expectations 
than a non-detection, and would likely provide strong evidence for a greater 
incidence of planets around young stars than MS stars. 

\subsubsection{Comparison with Other Cluster Surveys}

There have been many surveys searching for planetary transits in star clusters 
(see \citealt{weldrake2007} for a review). These surveys are unique relative 
to the many shallow, wide-field transit surveys, because they examine stars of 
a known age and metallicity. Therefore any planets or low mass EBs found in 
these clusters can be used as observational constraints on stellar 
environments and their evolution. 

These cluster surveys also present an opportunity to place upper limits on 
HJ incidence at a number of stellar ages. Unfortunately, few of these surveys 
have actually calculated upper limits on $f_P$, but we can compare our results 
with those that have. \citet{burke2006} surveyed the $\sim$1 Gyr cluster NGC 
1245, and found 95\% confidence upper limits for 1-3 day orbits which are 
a factor of $\sim$2 smaller than the upper limits from this work. They find 
upper limits of 6.4\% and 24\% for 1.5$R_{\rm Jup}$ and 1.0$R_{\rm Jup}$ HJs, 
respectively, compared to upper limits 
of 14\% and 52\%, respectively, from this work. We report an upper limit 
for 3-10 day period HJs of 45\%, which is lower than the \citet{burke2006} 
value of 52\% for 
1.5$R_{\rm Jup}$ planets. Neither study was able to place meaningful upper 
limits on the incidence of 1$R_{\rm Jup}$ planets in 3-10 day orbits. 
\citet{bramich2006} were able to place better upper limits on $f_p$ than those 
found in this study, however, they included both field and cluster stars 
($N_* \sim 30 000)$ in their analysis. 
\citet{weldrake2007b} observed 
31 000 stars in the globular cluster $\omega$ Centauri for 25 nights. These 
observations allowed them place a 95\% confidence level upper limit of 0.1\% 
on 1.5$R_{\rm Jup}$ planets in 1-3 day orbits. Given the significantly larger 
number of targets in both the \citet{bramich2006} and \citet{weldrake2007b} 
surveys it is unsurprising that they place more restricting upper limits on 
$f_p$.

\subsubsection {Extensions to the Remainder of Monitor}

We can also explore what happens when we extrapolate our results to Monitor as 
a whole. Monitor will observe a total of nearly 15 000 young ($<$ 200 Myr) 
stars, most of which have not yet reached the MS. This significant increase 
over the number of targets in NGC 2362 will lead to large reductions in the 
limits on $f_p$ for young stars. 

In fact, observations of an additional Monitor target, $h$ \& $\chi$ Per, 
which has $\sim$7500  cluster members \citep{aigrain2007}, will
lead to a considerable reduction in the limits on $f_p$ for young stars. 
Assuming the only difference between clusters is the number of 
observed stars\footnote{We note that this assumption is a significant over 
simplification of the actual situation given that the clusters are at 
different distances, are being observed with different telescopes, have 
different noise properties, and are different ages, however, we proceed simply 
to provide an order of magnitude estimate for the improvement in $f_p$.}, 
then we would be able to reduce all the upper limits 
in Table~\ref{table:upper-limit} by a factor of $\sim$13 following the 
observation of $h$ \& $\chi$ Per. When we consider all the stars to be 
observed by Monitor we will be able to reduce the limits on $f_p$ by a factor 
of $\sim$26. For the case of 1.5$R_{\rm Jup}$ planets in 3-10 day orbits this 
would mean an upper limit of $\sim$2.7\%.

\begin {table}%{lllrrrrrrlrrrrr}
\centering
\begin {tabular}{ccc}
   \hline
   \hline
$R_p$ & per & $f_p$   \\
($R_{\rm Jup}$) & (days) & (${\cal S}$ = ${\cal S}_{2362}$) \\
   \hline
1.0 & 1-3 &  0.031 \\
1.0 & 3-10 & 0.183  \\
1.5 & 1-3 &  0.008 \\
1.5 & 3-10 & 0.027  \\
   \hline
  \end {tabular}
\caption{Summary of the expected $\alpha$ = 0.01 upper limits on the fraction 
  of young stars with HJs following $\sim$100 hrs of observations of each of 
  the Monitor targets. $R_p$ and per are 
  the planet radius and orbital period, respectively. $f_p$ is the upper 
  limit on the fraction of stars with HJs assuming that 
  we achieve a sensitivity in all other clusters equal to that which we 
  achieved in NGC 2362. 
\label{table:n-stars}}
\end {table}

The above predictions assume that our achieved sensitivity in NGC 2362 will be 
the same in each of the clusters we have observed. Most of the Monitor targets 
are both older and less distant than NGC 2362, meaning in these cases we 
are probing stars with smaller radii and less intrinsic variability. As seen 
in Section 5.1.3, smaller radii should result in an 
improved sensitivity. Assuming a sensitivity equal to that in NGC 2362 we can 
estimate the final upper limits on $f_p$ we would expect from Monitor.
These values are summarised in Table~\ref{table:n-stars}. 

\subsubsection {Future Observational Considerations}

Finally, we would like to discuss the limitations of our survey of NGC 2362. 
We have learned that our sensitivity is largely limited by intrinsic 
stellar variability and stars with large radii. 
Both of these limitations could be reduced by observing older clusters: there 
will be less intrinsic variability while the stellar radii will be smaller 
given that the stars  have had more time to contract toward their MS radius. 

Another way to increase the sensitivity of our survey would be to decrease 
the point-to-point rms and the noise over 
time-scales equal to or longer than the duration of a transit. Given that the 
red and white noise properties of our data are dependent on a number of 
factors, including the detectors, sky noise, and observing conditions to name 
a few, there is no simple panacea for reducing the noise. However, there is 
one slight change in observing strategy which would generally reduce the noise 
(both red, which would have the greatest effect on our brightest targets, 
and white, which would have the greatest effect on our faintest targets) 
in the light curves: conducting all observations within a 
single observing season. Even in many of the light curves with low $\chi^2$ 
with respect to a flat model we notice small fluctuations, of order 0.01 mag, 
in the median flux following large gaps in our observations. These shifts are 
real and not the result of systematics. Making all the observations in a 
single observing season will remove this slight source of variability from 
our data and create a greater sensitivity to transits. We also note that  
this change in strategy would have the added benefit of significantly reducing 
the computing time necessary to adequately search the light curves for 
transits using our search algorithm.

\section {Summary and Conclusions}

We have conducted a search for occultations in NGC 2362. We observed the 
cluster on 18 nights from February 2005 to January 2006 with the Mosaic II 
imager on the 4m Blanco telescope at CTIO. We achieved an average cadence of 
$\sim$6 minutes, which is sufficient to perform differential photometry 
and search the data set for any transits. We used a $V, V-I$ CMD to 
photometrically select 1813 candidate cluster members.

Following the selection of candidate cluster members, we developed a 
systematic 
method for searching and identifying occultations in our light curves. This 
method consisted of two major steps: (1) the identification and removal of 
intrinsic stellar variability due to rotation, and (2) the search for transits 
using the occultation search algorithm of \citet{ai2004} modified to account 
for red noise. Following the 
removal of saturated stars and stars too faint for spectroscopic follow-up, 
we searched a total of 1180 stars for transits, of which about $\sim$475 are 
expected to be cluster members according to our contamination estimate. 

Our search identified 15 light curves with reductions in flux that passed all 
of our detection criteria. Only six of these systems, however, have observed 
variability that would be compatible with a planetary companion based on the 
observed occultation depth. Some of these systems may be cluster 
EBs, which would help to provide important constraints on the mass and 
radius, and by extension evolution, of PMS stars. 

Using a series of Monte Carlo simulations we predict the number of detectable 
HJs and find it is consistent with our null result. With 99\% confidence we 
place a limit on the fraction of stars in NGC 2362 with 1-3 d period HJs at
$<$22\%, while we limit the fraction with 3-10 d period HJs at $<$70\%, 
assuming a planetary radius of $1.5R_{\rm Jup}$. We compare these limits with 
other cluster surveys and find that we are not as sensitive to 
transits as other surveys, because we have far fewer targets. The limits 
for NGC 2362 provide 
observational constraints on the fraction of stars with HJs at an age $<$ 10 
Myr. From the simulations we also know that our sensitivity to transits 
increases as the stellar radius decreases, a somewhat non-intuitive result, 
which supports the findings of \citet{ap2007}.

Finally, we examine the 
prospects of the Monitor project as a whole. If we assume the same 
sensitivity is achieved for 
the entire Monitor survey, we will be able to place an upper limit on the 
number of young stars with a 1.5$R_{\rm Jup}$ HJ at $\sim$2.7\%, assuming a 
null detection.

\subsection*{Acknowledgments}

Based on observations obtained at CTIO, a division of the National Optical 
Astronomy Observatories, which is operated by the Association of Universities 
for Research in Astronomy, Inc. under cooperative agreement with the National 
Science Foundation. This publication makes use of data products from the 2MASS,
 which is a joint project of the University of Massachusetts and the Infrared 
Processing and Analysis Center/California Institute of Technology, funded by 
the National Aeronautics and Space Administration and the National Science 
Foundation. This research has also made use of the SIMBAD data base, operated 
at CDS, Strasbourg, France.

We wish to thank the referee for the thoughtful comments towards improving 
this paper. AM gratefully acknowledges the support of a Gates-Cambridge Trust 
fellowship. AM would like to thank Cathie Clarke and Keivan Stassun for their 
useful comments regarding the content of this paper. JI gratefully 
acknowledges the support of a PPARC studentship, and SA the support of a 
PPARC postdoctoral fellowship.

\label{lastpage}

\end{document}